\documentclass[preprint,showpacs,preprintnumbers,amsmath,amssymb]{revtex4}

\usepackage{graphicx}
\usepackage{dcolumn}
\usepackage{bm}
\usepackage{amssymb}

\begin{document}
\newcommand{\vecx}{\mbox{\boldmath $x$}}
\newcommand{\vecp}{\mbox{\boldmath $p$}}

\title{
Thermal entanglement of Hubbard dimers \\
in the nonextensive statistics
}

\author{Hideo Hasegawa
\footnote{hideohasegawa@goo.jp}
}
\affiliation{Department of Physics, Tokyo Gakugei University,  
Koganei, Tokyo 184-8501, Japan}%

\date{\today}

\begin{abstract}
The thermal entanglement of the Hubbard dimer (two-site Hubbard model)
has been studied with the nonextensive statistics.
We have calculated the auto-correlation ($O_q$), 
pair correlation ($L_q$), concurrence ($\Gamma_q$)
and conditional entropy ($R_q$)
as functions of entropic index $q$ and the temperature $T$.
The thermal entanglement is shown to considerably
depend on the entropic index. For $q < 1.0$,
the threshold temperature where $\Gamma_q$
vanishes or $R_q$ changes its sign is more increased
and the entanglement may survive at higher temperatures than for $q=1.0$.  
Relations among $L_q$, $\Gamma_q$ and $R_q$ are investigated.
The physical meaning of the entropic index $q$ is discussed
with the microcanonical and superstatistical approaches.
The nonextensive statistics is applied also to Heisenberg dimers.

\end{abstract}




\maketitle
\newpage

\section{INTRODUCTION}


Quantum entanglement is one of the most intriguing subjects
in quantum information theory, and it has been investigated
from various viewpoints in the last decade 
(for a review, see Refs. \cite{QI,Amico08}, related references therein).
Quantum entanglement is expected to play an essential role as a resource 
in quantum communication and information processing.
Many studies have been made on quantum entanglement with
quantum spin models \cite{Arnesen01}-\cite{Chan06}
and fermion systems \cite{Scliemann01b}-\cite{Coe10}
to clarify both its qualitative and quantitative aspects.
The interface between the quantum information
and statistical mechanics has been considerably investigated
in recent years. It has been shown that
entanglement of two neighboring sites shows a sharp peak
either near or at the critical point where
the phase transition takes place 
\cite{Osterloh02,Osborne02,Vidal03,Gu04}. 
This suggests an intimate
relation between quantum entanglement and quantum phase 
transition \cite{Amico08,Larsson06,Anfossi07,Wu06,Carr10}.

These studies mentioned above have been made within the Boltzmann-Gibbs statistics (BGS).
In the last several years,  
there is an increased interest in the
nonextensive statistics (NES), which was initially
proposed by Tsallis \cite{Tsallis88,Curado91,Tsallis98}.
This is because the standard method based on
the BGS cannot properly deal with
{\it nonextensive} systems where the physical quantities such as
the energy and/or entropy of the $N$-body system are not
proportional to $N$.
The nonextensivity has been realized in the three typical 
systems: (1) systems with long-range interactions,
(2) small-scale systems with fluctuations of temperatures
or energy dissipations, and (3) multi-fractal systems \cite{Tsallis04}.
For example, in a gravitating system with the long-range interaction,
which is a typical case (1), the specific heat 
becomes negative \cite{Padman90}. 
A cluster of 147 sodium atoms, which belongs to the case (2), 
has been reported to show
the negative specific heat \cite{Schmidt01}.
The generalized entropy (referred to as the Tsallis entropy)
proposed by Tsallis \cite{Tsallis88}\cite{Tsallis98} 
is given by, with ${\rm Tr}\:\hat{\rho} =1$:
\begin{equation}
S_q= k \left( \frac{{\rm Tr} \:\hat{\rho}^q-1}{1-q} \right),
\label{eq:A1}
\end{equation}
where $k$ is a positive constant, $\hat{\rho}$ the density matrix,
Tr the trace and $q$ the entropic index. 
In the limit of $q \rightarrow 1$, Eq. (\ref{eq:A1}) reduces to
the von Neumann form,
\begin{equation}
S_1 = - k \:{\rm Tr}\: \hat{\rho} \ln \hat{\rho},
\label{eq:A2}
\end{equation}
where ${\rm Tr}\:\hat{\rho} =1$.
We will set $k=k_B$ ($k_B$: the Boltzmann constant)
when we discuss the thermodynamical entropy.
The nonextensivity (non-additivity) in the Tsallis entropy 
is explained as follows. For a system consisting of two  
independent subsystems A and B with the density matrices,
$\hat{\rho}(A)$ and $\hat{\rho}(B)$,
the density matrix of the total system is described by
$\hat{\rho}(A,B)=\hat{\rho}(A) \otimes \hat{\rho}(B)$
from which the Tsallis entropy is given by
\begin{equation}
S_q(A,B)=  S_q(A)+S_q(B)+\frac{(1-q)}{k_B} S_q(A) S_q(B),
\label{eq:A3}
\end{equation}
$S_q(\eta)$ denoting the entropy of the subsystem $\eta$ ($=A, B$). 
Eq. (\ref{eq:A3}) shows that the entropy is additive
for $q=1$ and non-additive for $q \neq 1$.
The quantity $(q-1)$ expresses the measure of
the nonextensivity.
The NES has been applied to a wide class of subjects such as
physics, chemistry, and biology \cite{Tsallis09,NES}. 

Small-scale systems belong to nonextensive systems as mentioned above.
It is necessary to take into account the effect of
the nonextensivity, when we study the properties of quantum information
of qubits, which have been mainly investigated within the BGS
\cite{Scliemann01b}-\cite{Coe10}.
In recent years, the NES has been applied to a study on quantum information
\cite{Rajagopal99}-\cite{Rajagopal05}.
It has been shown that the non-additive Tsallis entropy
yields a better measure for the separability criterion of entanglement
than the additive von Neumann entropy \cite{Rajagopal99,Abe99,Abe01,Rajagopal05}.
Ref. \cite{Barranco99} discusses an NES generalization of the von Neumann mutual 
information which is shown to strongly depend on the entropic index $q$.
The above list of applications of the NES to
quantum information is not complete: relevant references
are presented in \cite{Tsallis09,NES,Rajagopal05}.

It is worthwhile to apply the NES to the Hubbard model \cite{Hubbard64}, 
which is a typical model for strongly correlated fermion systems and
which has been widely adopted for a study 
on quantum information
\cite{Scliemann01b}-\cite{Coe10}. 
Despite the simplicity of the model, however, it is very difficult to
obtain its exact solution except for some limited cases. 
In order to obtain a reasonable solution, various approximate
methods have been proposed and developed.
The Hubbard model provides us with good
qualitative description for many interesting phenomena
such as electron correlation, magnetism, superconductivity
and quantum entanglement.
We may employ a finite-size Hubbard model
to obtain an analytical, exact solution.
The Hubbard dimer (two-site Hubbard model)
has been adopted as a simple model which can be analytically solved.
Although the Hubbard dimer seems a toy model,
it has played an important role as a model of qubits
in the theory of quantum information 
\cite{Gu04,Deng04,Dowling06}.  

In our previous papers \cite{Hasegawa04,Hasegawa05,Hasegawa06,Hasegawa07}, 
we applied the NES to Hubbard dimers
to investigate effects of the nonextensivity on their thermodynamical and
magnetic properties, bearing small-size systems in mind. 
It is interesting to examine the effect of the nonextensivity on
the properties of the quantum entanglement of two qubits  \cite{Amico08}
described by the Hubbard dimer within the NES, which is the main purpose
of the present paper. 
Among various quantities expressing thermal entanglement, we have calculated 
the pair correlation, concurrence \cite{Hill97,Wootters98}
and conditional  entropy \cite{Cerf97,Cerf99}. 
As will be shown in our study, the entropic index $q$ has considerable
effects on the properties of thermal entanglement 
which may be improved by the nonextensivity.
The concurrence of the Hubbard dimer has been discussed
within the BGS \cite{Gu04,Deng04,Dowling06}.
The generalization of the conditional entropy to the NES has been proposed
in Refs. \cite{Abe01,Abe02}.
This is the first systematic study on the thermal entanglement 
of nonextensive fermion systems as far as we are aware of.

The paper is organized as follows. In Sec. 2, after briefly reviewing
the maximum-entropy method (MEM) in the NES \cite{Tsallis88}, 
we derive the density matrix to obtain 
auto- and pair correlations, concurrence and conditional entropy.
In Sec. 3 we present model calculations of relevant quantities
as functions of the entropic index $q$ and the temperature $T$.
In Sec. 4, we make a comparison among 
the pair correlation, concurrence and conditional entropy.
Effects of magnetic field and interatomic interactions 
in the adopted model are investigated.
The physical meaning of the entropic index $q$ is discussed
with the use of the microcanonical approach (MCA) 
\cite{Plastino94,Almeida01,Potiguar03,Aringazin03,Johal03}
and superstatistical approach (SSA) \cite{Wilk00,Beck02,Cohen04}.
Sec. 5 is devoted to our conclusion.

\section{Formulations}

\subsection{Hubbard dimers}

We consider the extended Hubbard dimer whose Hamiltonian
is given by
\begin{eqnarray}
\hat{H} &=& -t \sum_{\sigma}
( a_{1\sigma}^{\dagger} a_{2\sigma} 
+  a_{2\sigma}^{\dagger} a_{1\sigma}) 
+ U \sum_{j=1}^2 n_{j \uparrow} n_{j \downarrow } 
+ V_1 \sum_{\sigma}n_{1\sigma} n_{2\sigma}
+ V_2 \sum_{\sigma}n_{1\sigma} n_{2 -\sigma} \nonumber \\
&-& \mu_B B \sum_{j=1}^2 (n_{j \uparrow} - n_{j \downarrow}), 
\label{eq:B1} 
\end{eqnarray}
where 
$n_{j\sigma}= a_{j\sigma}^{\dagger} a_{j\sigma}$,
$a_{j\sigma}$ denotes the annihilation operator of an electron with
spin $\sigma$ ($=\uparrow, \downarrow $) on a site $j$ (=1, 2), 
$t$ the hopping integral,
$U$ the intraatomic interaction between electrons with opposite spin, 
$V_1$ ($V_2$) the interatomic Coulomb interaction
between the same (opposite) spin, $\mu_B$ the Bohr magneton 
and $B$ an applied magnetic field.
By using the standard basis for half-filled case with two electrons 
given by
\begin{eqnarray}
\vert \Psi_1 \rangle &=& \vert \uparrow\downarrow \rangle_1 \vert 0 \rangle_2,\;
\vert \Psi_2 \rangle = \vert 0\rangle_1 \vert \uparrow\downarrow\rangle_2 ,\;
\vert \Psi_3 \rangle = \vert \uparrow\rangle_1 \vert \downarrow\rangle_2, \nonumber \\
\vert \Psi_4 \rangle &=& \vert \downarrow\rangle_1 \vert \uparrow\rangle_2,\;\;\;
\vert \Psi_5 \rangle = \vert \uparrow\rangle_1 \vert \uparrow\rangle_2,\;\;
\vert \Psi_6 \rangle = \vert \downarrow \rangle_1 \vert \downarrow\rangle_2,
\nonumber 
\end{eqnarray}
we obtain the Hamiltonian matrix,
\[
H=\left(
\begin{array}{cccccc}
\;\;U \;\;& \;\; 0 \;\;& \;\;-t \;\; & \;\;-t \;\;&\;\; 0 \;\;& \;\;0\;\; \\
0    &  U   & -t  & -t   & 0   & 0 \\
\;\;-t \;\;  & \;\;-t \;\;  & V_2 &  0   & 0   & 0 \\
-t   & -t   &  0  & V_2  & 0   & 0 \\
0    &  0   &  0  &  0   & V_1-2 \mu_B B & 0  \\
0    &  0   &  0  &  0   & 0 & V_1+2 \mu_B B   
\end{array}
\right).\label{eq:B3}
\]
Six eigenvalues of the system are given by \cite{Deng04,Navarro09}
\begin{eqnarray}
\epsilon_i &=& \frac{1}{2}(U+V_2-D),
\;\frac{1}{2}(U+V_2+D),\;U,\;V_2, \;V_1-2 \mu_B B,\;V_1+2 \mu_B B \nonumber \\
&&\hspace{10cm}\mbox{for $i=1$ to $6$},
\label{eq:B4}
\end{eqnarray}
and the corresponding eigenvectors are given by
\begin{eqnarray}
\vert \Phi_1\rangle &=& \frac{\sin \theta}{\sqrt{2}}
\left( \vert \uparrow \downarrow \rangle_1 \vert 0 \rangle_2
+\vert 0 \rangle_1 \vert \uparrow \downarrow\rangle_2 \right)
+\frac{\cos \theta}{\sqrt{2}}
\left( \vert \uparrow \rangle_1 \vert \downarrow \rangle_2
+\vert \downarrow \rangle_1 \vert \uparrow \rangle_2 \right), \\
\vert \Phi_2 \rangle &=& \frac{\cos \theta}{\sqrt{2}}
\left( \vert \uparrow \downarrow \rangle_1 \vert 0 \rangle_2
+\vert 0 \rangle_1 \vert \uparrow \downarrow \rangle_2 \right)
-\frac{\sin \theta}{\sqrt{2}}
\left( \vert \uparrow \rangle_1 \vert \downarrow \rangle_2
+\vert \downarrow \rangle_1 \vert \uparrow \rangle_2 \right), \\
\vert \Phi_3 \rangle &=& \frac{1}{\sqrt{2}}
\left( \vert \uparrow \downarrow \rangle_1 \vert 0 \rangle_2 
-\vert 0 \rangle_1 \vert \uparrow \downarrow\rangle_2 \right),\\
\vert \Phi_4 \rangle &=& \frac{1}{\sqrt{2}}
\left( \vert \uparrow \rangle_1 \vert \downarrow \rangle_2 
- \vert \downarrow \rangle_1 \vert \uparrow \rangle_2 \right), \\
\vert \Phi_5 \rangle &=& \vert \uparrow \rangle_1 \vert \uparrow \rangle_2, \\
\vert \Phi_6 \rangle &=& \vert \downarrow \rangle_1 \vert \downarrow \rangle_2,
\end{eqnarray}
where
\begin{eqnarray}
\tan \theta &=& \frac{4 t}{U-V_2+D},\\
D &=& \sqrt{(U-V_2)^2+16 t^2}.
\end{eqnarray}
For $t/U \ll 1$ with $V_1=V_2=B=0$, we obtain
\begin{eqnarray}
\epsilon_1 &=& - \frac{4t^2}{U},\;\epsilon_2=U+\frac{4t^2}{U},
\;\epsilon_3=U,\;\epsilon_4=\epsilon_5=\epsilon_6=0, 
\label{eq:K1}\\
\sin^2 \theta &=& \frac{4t^2}{U^2}, \;\; \cos^2 \theta = 1-\frac{4t^2}{U^2},
\label{eq:K2}
\end{eqnarray}
where $\epsilon_1$ is the lowest eigenstate for $U > 0$.

The partition function in the BGS is given by 
\begin{eqnarray}
Z_{BG} &=& Z_1 = {\rm Tr} \:e^{-\beta \:H}=\sum_i \:e^{-\beta \epsilon_i} \nonumber\\
&=& 2 \:e^{-\beta(U+V_2)/2} \cosh\left( \frac{\beta D}{2}\right)
+ e^{-\beta U} + e^{-\beta V_2}
+ 2\: e^{-\beta V_1} \cosh(2 \beta \mu_B B),
\label{eq:B5}
\end{eqnarray}
where $\beta=1/k_B T$.
From Eq. (\ref{eq:B5}) we can obtain various thermodynamical 
quantities of the system. 

\subsection{Maximum-entropy method in the NES}

We will study the Hubbard dimer given by Eq. (\ref{eq:B1}) 
within the NES, where the PDF
or the density matrix is evaluated by the MEM for the Tsallis entropy.
At the moment there are four MEMs in the NES:
(a) original method \cite{Tsallis88},
(b) un-normalized method \cite{Curado91}, 
(c) normalized method \cite{Tsallis98}, and 
(d) optimal Lagrange multiplier (OLM) method \cite{Martinez00}.
The four MEMs are compared in Ref. \cite{Tsallis04}.
Among the four MEMs, (c) normalized MEM and (d) OLM-MEM
with the $q$-average
have been widely adopted for a study of nonextensive systems,
because they are believed to be more superior than (a) original MEM \cite{Tsallis88}
and (b) un-normalized MEM \cite{Curado91,Tsallis04}.
Recent papers \cite{Abe09,Abe09b,Lutsko09}, however, have pointed out
that thermodynamical quantities obtained by the $q$-average 
are unstable for a small change in the PDF, 
whereas those obtained by the normal average are stable \cite{Hanel09}.
The stability of the $q$-average is currently controversial
\cite{Abe09}-\cite{Hasegawa10b}. 
Although (c) normalized MEM \cite{Tsallis98} with the $q$-average
was adopted in our previous papers 
\cite{Hasegawa04,Hasegawa05,Hasegawa06,Hasegawa07}\cite{Note1},
we have employed in the present study, (a) original MEM 
with the normal average \cite{Tsallis88,Abe09}.
In Appendix A, thermodynamical quantities of the entropy, 
specific heat and susceptibility calculated by (a) original MEM \cite{Tsallis88,Abe09}
are summarized and compared to
previous ones obtained by (c) normalized MEM \cite{Tsallis98} with the $q$-average
\cite{Hasegawa04,Hasegawa05,Hasegawa06,Hasegawa07}\cite{Note1}.
In Appendix B the NES with (a) original MEM \cite{Tsallis88,Abe09}
is applied also to Heisenberg dimers.

Imposing the two constraints given by 
\begin{eqnarray}
{\rm Tr} \:(\hat{\rho}_q)&=&1, 
\label{eq:C1}\\
{\rm Tr} \:(\hat{\rho}_q \;\hat{H}) &=& \langle \hat{H}\rangle_q
=E_q,
\label{eq:C2}
\end{eqnarray}
we obtain the density matrix given by
\begin{equation}
\hat{\rho}_q=\frac{1}{X_q} 
{\rm Exp}_q[-\beta(H-E_q)],
\label{eq:C4}
\end{equation}
with 
\begin{eqnarray}
X_q &=& {\rm Tr} \:{\rm Exp}_q[-\beta(H-E_q)],
\label{eq:C5}
\end{eqnarray}
where $\langle \cdot \rangle_q$ expresses the average over 
$\hat{\rho}_q$, $\beta$ the inverse of the temperature
and ${\rm Exp}_q(x)$ is defined by \cite{Abe09}
\begin{eqnarray}
{\rm Exp}_q(x) &=& [1+(1-1/q)x]_+^{1/(q-1)},
\label{eq:C6}
\end{eqnarray}
with $[y]_+={\rm max}(y,0)$.
Note that ${\rm Exp}_q(x)$ is different from the conventional
$q$-exponential function $\exp_q(x)$ defined by \cite{Tsallis88} 
\begin{eqnarray}
\exp_q(x) &=& [1+(1-q)x]_+^{1/(1-q)}.
\label{eq:Y6}
\end{eqnarray}
The two $q$-exponential functions, ${\rm Exp}_q(x)$ and $\exp_q(x)$,
have the relation \cite{Abe09}:
\begin{eqnarray}
\exp_q(x) &=& {\rm Exp}_{2-q}((2-q)x), \;\;
{\rm Exp}_q(x) = \exp_{2-q}(x/q).
\label{eq:Y7}
\end{eqnarray}
Both ${\rm Exp}_q(x)$ and $\exp_q(x)$ reduce to 
the exponential function $\exp(x)$ in the limit of $q \rightarrow 1.0$.

\subsection{Auto- and pair correlations}

For the Hubbard dimer under consideration, we obtain
\begin{eqnarray}
\hat{\rho}_q &=& \frac{1}{X_q}\sum_i w_i\: 
\vert \Phi_i \rangle \langle \Phi_i \vert, 
\label{eq:C4b}\\
E_q &=& \frac{1}{X_q} \sum_i \:w_i \:\epsilon_i, 
\label{eq:D1}\\
X_q &=& \sum_i \:w_i, 
\label{eq:D2}
\end{eqnarray}
where
\begin{eqnarray}
w_i &=& {\rm Exp}_q[-\beta(\epsilon_i -E_q) ].
\label{eq:D3}
\end{eqnarray}
The energy $E_q$ in Eq. (\ref{eq:D1}) includes the partition function
$X_q$ which is expressed in terms of
$E_q$ in Eq. (\ref{eq:D3}). Then $E_q$ and $X_q$ are self-consistently determined 
by Eqs. (\ref{eq:D1})-(\ref{eq:D3}) for given $q$ and $\beta$.

We first consider auto- ($O_q$) and pair correlations ($L_q$) defined by
\begin{eqnarray}
O_q&=& 1-  \sum_{j=1}^2 \left< n_{j \uparrow} n_{j \downarrow } \right>_q, 
\label{eq:E1}\\
L_q &=&   \sum_{\sigma} \left< n_{1\sigma} n_{2\sigma}
-n_{1\sigma} n_{2 -\sigma} \right>_q.
\label{eq:E2} 
\end{eqnarray}
When we employ the relations given by
\begin{eqnarray}
\sum_{j=1}^2 \left< n_{j \uparrow} n_{j \downarrow } \right>_q
&=&\left< \frac{\partial H}{\partial U}\right>_q, 
\label{eq:E13}\\
\sum_{\sigma} \left<  n_{1\sigma} n_{2\sigma} \right>_q
&=& \left< \frac{\partial H}{\partial V_1} \right>_q, 
\label{eq:E14}\\
\sum_{\sigma} \left< n_{1\sigma} n_{2 -\sigma} \right>_q
&=& \left<\frac{\partial H}{\partial V_2} \right>_q,
\label{eq:E15}
\end{eqnarray}
Eqs. (\ref{eq:E1}) and (\ref{eq:E2}) become
\begin{eqnarray}
O_q 
&=& 1 - \left< \frac{\partial H}{\partial U}\right>_q, 
\label{eq:E1b}\\
L_q 
&=& \left< \frac{\partial H}{\partial V_1} \right>_q
- \left<\frac{\partial H}{\partial V_2} \right>_q.
\label{eq:E2b} 
\end{eqnarray}

We may evaluate $\left< \partial H/\partial \theta \right>_q $
with $\theta=U$, $V_1$, and $V_2$ as follows.
Taking the derivative of $X_q$ in Eq. (\ref{eq:C5}) with respect to $\theta$,
we obtain
\begin{eqnarray}
\frac{\partial X_q}{\partial \theta}
&=& -\beta \:{\rm Tr} \{ \left( {\rm Exp}_q[-\beta (H-E_q)]\right)
\left( \frac{\partial H}{\partial \theta}
-\frac{\partial E_q}{\partial \theta}  \right) \},\\
&=& -\beta X_q \:\left( \left<\frac{\partial H}{\partial \theta}\right>_q
-\frac{\partial E_q}{\partial \theta}  \right),
\end{eqnarray}
which leads to 
\begin{eqnarray}
\left< \frac{\partial H}{\partial \theta}\right>_q
&=& \frac{\partial E_q}{\partial \theta}
-\frac{1}{\beta X_q} \frac{\partial X_q}{\partial \theta}.
\label{eq:E4}
\end{eqnarray}
From Eqs. (\ref{eq:D1})-(\ref{eq:D3}),
self-consistent equations for $\partial E_q/\partial \theta$
and $\partial X_q/\partial \theta$ are given by
\begin{eqnarray}
\frac{\partial E_q}{\partial \theta} 
&=& a_{11} \frac{\partial E_q}{\partial \theta} 
+ a_{12} \frac{\partial X_q}{\partial \beta} + c_{1 \theta}, 
\label{eq:E5}\\
\frac{\partial X_q}{\partial \theta} 
&=& a_{21} \frac{\partial E_q}{\partial \theta}
+ a_{22} \frac{\partial X_q}{\partial \beta}+c_{2 \theta},
\label{eq:E6}  
\end{eqnarray}
with
\begin{eqnarray}
%
c_{2 \theta} &=& -\beta \sum w_i
\:\left( \frac{\partial \epsilon_i}{\partial \theta} \right),
\label{eq:E7}
\end{eqnarray}
where an explicit expression for $c_{1 \theta}$ is not necessary (see below).
Solving Eqs. (\ref{eq:E5})-(\ref{eq:E7}) with respect to
$\partial E_q/\partial \theta$ and $\partial X_q/\partial \theta$
and substituting them to Eq. (\ref{eq:E4}), we obtain
\begin{eqnarray}
\left< \frac{\partial H}{\partial \theta }\right>_q
&=& -\frac{c_{2\theta}}{a_{21}}
= \frac{1}{X_q} \sum_i w_i\:
\left( \frac{\partial \epsilon_i}{\partial \theta} \right).
\label{eq:E8}
\end{eqnarray}
Simple calculations with the use of Eqs. (\ref{eq:B4}) and (\ref{eq:E8}) lead to
\begin{eqnarray}
\left< \frac{\partial H}{\partial U}\right>_q
&=& \frac{1}{X_q}\left[\frac{1}{2}\left(1-\frac{U-V_2}{D} \right) \:w_1
+ \frac{1}{2}\left(1+\frac{U-V_2}{D} \right) \:w_2+w_3 \right], 
\label{eq:E9}\\
\left< \frac{\partial H}{\partial V_1}\right>_q
&=& \frac{1}{X_q}(w_5+w_6), 
\label{eq:E10}\\
\left< \frac{\partial H}{\partial V_2}\right>_q
&=& \frac{1}{X_q}\left[\frac{1}{2}\left(1+\frac{U-V_2}{D} \right) \:w_1
+ \frac{1}{2}\left(1-\frac{U-V_2}{D} \right) \:w_2+w_4 \right].
\label{eq:E11}
\end{eqnarray}
Substituting Eqs. (\ref{eq:E9})-(\ref{eq:E11}) to Eqs. (\ref{eq:E1b}) 
and (\ref{eq:E2b}), we finally obtain
\begin{eqnarray}
O_q &=& 1- \frac{1}{X_q}\left[\frac{1}{2}\left(1-\frac{U-V_2}{D} \right) \:w_1
+ \frac{1}{2}\left(1+\frac{U-V_2}{D} \right) \:w_2+w_3 \right], 
\label{eq:E12}\\
L_q &=&
\frac{1}{X_q}\left[w_5+w_6
-\frac{1}{2}\left(1+\frac{U-V_2}{D} \right) \:w_1
- \frac{1}{2}\left(1-\frac{U-V_2}{D} \right) \:w_2-w_4 \right].
\label{eq:E12b}
\end{eqnarray}

In the limit of $q \rightarrow 1$, Eqs. (\ref{eq:E12}) and (\ref{eq:E12b}) 
reduce to
\begin{eqnarray}
O_1 &=& 1- \frac{1}{Z_1}\left[\frac{1}{2}\left(1-\frac{U-V_2}{D} \right) 
\:e^{-\beta \epsilon_1}
+ \frac{1}{2}\left(1+\frac{U-V_2}{D} \right) \:\:e^{-\beta \epsilon_2}
+\:e^{-\beta \epsilon_3} \right], \\
L_1 &=&
\frac{1}{Z_1}\left[\:e^{-\beta \epsilon_5}+\:e^{-\beta \epsilon_6}
-\frac{1}{2}\left(1+\frac{U-V_2}{D} \right) \:e^{-\beta \epsilon_1}
- \frac{1}{2}\left(1-\frac{U-V_2}{D} \right) \:e^{-\beta \epsilon_2}
-e^{-\beta \epsilon_4} \right]. \nonumber \\
&&
\end{eqnarray}

In the limit of $T=0$, the auto-correlation in the BGS and NES is given by
\begin{eqnarray}
O_q &=& - L_q = \frac{1}{2}\left(1+\frac{U-V_2}{D} \right), 
\label{eq:E16}\\
&=& \left\{ \begin{array}{ll}
\frac{1}{2}
\quad & \mbox{for $(U-V_2)/t=0$}, \\
1
\quad & \mbox{for $(U-V_2)/t \rightarrow \infty$}. 
\end{array} \right.
\label{eq:E17}
\end{eqnarray}

\subsection{Concurrence}

The concurrence $\Gamma$ has been proposed as a measure of entanglement
for systems of two qubits \cite{Hill97,Wootters98}.
It is defined with eigenvalues $\lambda_1^2 \geq \cdot \cdot \geq \lambda_4^2$
for the positive Hermitean matrix 
$\hat{R}=\sqrt{\rho} \tilde{\rho} \sqrt{\rho}$ by
\cite{Hill97,Wootters98}
\begin{eqnarray}
\Gamma &=& \max (\lambda_1-\lambda_2-\lambda_3-\lambda_4, 0),
\end{eqnarray}
where $\tilde{\rho}=(\sigma^y \otimes \sigma^y) \rho^* (\sigma^y \otimes \sigma^y)$
and $^*$ denotes the complex conjugate. 
The entanglement of formation $E_F$ \cite{Bennet96} is expressed
in terms of $\Gamma$ by 
\begin{eqnarray}
E_F &=& - \sum_{\xi=\pm}
\left( \frac{1+\xi \sqrt{1- \:\Gamma^2}}{2} \right) 
\log_2 \left( \:\frac{1+\xi \sqrt{1 -\Gamma^2}}{2} \right).
\end{eqnarray}
For a pair of qubits of the Hubbard dimer, the concurrence is given by 
\cite{Gu04,Deng04}
\begin{eqnarray}
\Gamma &=& \max \left(\vert  \sum_{\sigma}
\left< a_{1\sigma}^{\dagger} a_{2\sigma} \right> 
\vert
- \sum_{\sigma}\left<  n_{1\sigma} n_{2\sigma} \right>, 0\right), 
\end{eqnarray}
where the bracket $\langle \cdot \rangle$ means the average
over the density matrix.
It is straightforward to generalize the method of Ref. \cite{Deng04} to the NES,
in which the $q$-dependent concurrence $\Gamma_q$ is given by
\begin{eqnarray}
\Gamma_q &=& \max \left(\vert  \sum_{\sigma}
\left< a_{1\sigma}^{\dagger} a_{2\sigma} \right>_q 
\vert
- \sum_{\sigma}\left<  n_{1\sigma} n_{2\sigma} \right>_q, 0\right), \\
&=&\max \left( \frac{1}{2}\mid \left< \frac{\partial H}{\partial t} \right>_q\mid
- \left< \frac{\partial H}{\partial V_1} \right>_q, 0 \right).
\label{eq:F1}
\end{eqnarray}
In deriving Eq. (\ref{eq:F1}), we employ the relations given 
by Eq. (\ref{eq:E14}) and by
\begin{eqnarray}
\sum_{\sigma} \left< a_{1\sigma}^{\dagger} a_{2\sigma} 
+  a_{2\sigma}^{\dagger} a_{1\sigma} \right>_q
&=& - \left< \frac{\partial H}{\partial t} \right>_q.
\label{eq:F2}
\end{eqnarray}

By using Eqs. (\ref{eq:B4}) and (\ref{eq:E8}) with $\theta=t$,
we may calculate $\langle \partial H/\partial t \rangle_q$,
\begin{eqnarray}
\left< \frac{\partial H}{\partial t}\right>_q
&=& - \frac{8 t}{X_q D}(w_1-w_2). 
\label{eq:F3}
\end{eqnarray}
Substituting Eqs. (\ref{eq:E10}) and (\ref{eq:F3}) to Eq. (\ref{eq:F1}), 
we obtain
\begin{eqnarray}
\Gamma_q &=& \frac{1}{X_q}{\rm max} 
\left[ \left( \frac{4t}{D} \right) \vert \: (w_1-w_2) \: \vert-(w_5+w_6), 0
\right].
\label{eq:F4}
\end{eqnarray}

In the limit of $T=0$, $\Gamma_q$ in both the NES and BGS is given by
\begin{eqnarray}
\Gamma_q &=& \frac{4t}{D}, 
\label{eq:F6}\\
&=& \left\{ \begin{array}{ll}
1
\quad & \mbox{for $(U-V_2)/t=0$}, \\
0
\quad & \mbox{for $(U-V_2)/t \rightarrow \infty$}.
\end{array} \right.
\label{eq:F7}
\end{eqnarray}

In the limit of $q=1.0$, Eq. (\ref{eq:F4}) reduces to
\begin{eqnarray}
\Gamma_q &=& \frac{1}{Z_1}{\rm max} 
\left[ \left( \frac{4t}{D} \right) \vert \: 
(e^{-\beta \epsilon_1}- e^{-\beta \epsilon_2}) \: \vert
-(e^{-\beta \epsilon_5}+e^{-\beta \epsilon_6}), 0 \right],
\label{eq:F5}
\end{eqnarray}
which is in agreement with the result of Ref. \cite{Deng04}.
With increasing the temperature, the concurrence
is decreased and vanishes above the threshold temperature,
as will be shown in Sec. 3B.

\subsection{Conditional entropy}
The conditional entropy for subsystems A and B in the NES is expressed
by \cite{Abe01,Abe02}
\begin{eqnarray}
S_q(A \vert B) 
&=& \frac{S_q(A,B)-S_q(B)}{1+(1-q)S_q(B)/k_B},
\label{eq:J1}
\end{eqnarray}
with
\begin{eqnarray}
S_q(A,B) &=& k_B\:\frac{{\rm Tr} \: [\rho_q(A, B)]^q-1}{1-q}, \\
S_q(B) &=& k_B\: \frac{{\rm Tr}_B \: [\rho_q(B)]^q-1}{1-q}, \\
\hat{\rho}_q(B) &=& {\rm Tr}_A \:\hat{\rho}_q(A, B),
\label{eq:J2}
\end{eqnarray}
where ${\rm Tr}_A$ stands for the partial trace over the state $A$ 
and $\hat{\rho}_q(B)$ denotes the marginal density operator.
In the limit of $q \rightarrow 1$, $S_q(A \vert B)$ 
reduces to the von Neumann conditional entropy, 
$S_1(A \vert B)=S_1(A,B) -S_1(B)$,
whose properties have been discussed in Refs. \cite{Cerf97,Cerf99}.
In independent subsystems $A$ and $B$ where the relation:
$\hat{\rho}(A, B)=\hat{\rho}(A) \otimes \hat{\rho}(B)$ holds,
Eqs. (\ref{eq:A3}) and (\ref{eq:J1}) lead to 
$S_q(A \vert B)=S_q(A)$ \cite{Abe01,Abe02}.

Regarding subsystems $A$ and $B$ as sites 1 and 2 in the Hubbard dimer 
under consideration, we may obtain the marginal density operator given by
\begin{eqnarray}
\hat{\rho}_q(1) &=& {\rm Tr}_2 \:\hat{\rho}_q(1,2) \nonumber \\
&=&\frac{1}{X_q}
\left( g_1 \vert 0 \rangle_1 \langle 0 \vert_1
+g_2 \vert \uparrow \rangle_1 \langle \uparrow \vert_1
+ g_3 \vert \downarrow\rangle_1 \langle \downarrow \vert_1
+g_4 \vert \uparrow \downarrow \rangle_1 
\langle \uparrow \downarrow \vert_1 \right),
\label{eq:J3}
\end{eqnarray}
with
\begin{eqnarray}
g_1 &=& g_4=\frac{1}{2}\left( w_1 \sin^2 \theta +w_2 \cos^2 \theta 
+w_3 \right), \\
g_2 &=& \frac{1}{2}\left( w_1 \cos^2 \theta +w_2 \sin^2 \theta 
+w_4 + 2 w_5 \right), \\
g_3 &=& \frac{1}{2}\left( w_1 \cos^2 \theta +w_2 \sin^2 \theta 
+w_4 + 2 w_6 \right),
\label{eq:J4}
\end{eqnarray}
where $\hat{\rho}_q(1,2)=\hat{\rho}_q$ given by Eq. (\ref{eq:C4b}).
From Eqs. (\ref{eq:J1})-(\ref{eq:J4}), the conditional entropy is given by
\begin{eqnarray}
R_q &\equiv & S_q(1\vert 2)= S_q(2\vert 1)
= \frac{k_B}{(1-q)}\left( \frac{c_q}{d_q}-1 \right),
\label{eq:J5}
\end{eqnarray}
where
\begin{eqnarray}
c_q&=& Tr \:(\hat{\rho}_q^q)
= \frac{1}{X_q^q}\sum_i\: w_i^q = X_q^{1-q},
\label{eq:D5} \\
d_q &=& {\rm Tr}_1\:[\hat{\rho}_q(1)]^q
= \frac{1}{X_q^q}\left(g_1^q+g_2^q+g_3^q+g_4^q \right).
\label{eq:J6}
\end{eqnarray}
  
When $\epsilon_1$ is the lowest eigenstate at $T=0.0$,  
we obtain $w_1/X_q=1.0$ and $w_i/X_q=0.0$ for $i \neq 1$,
which lead to
\begin{eqnarray}
\hat{\rho}_q(1,2) &=& \vert \Phi_1 \rangle \langle \Phi_1 \vert, \\
\hat{\rho}_q(1) &=& \left( \frac{\cos^2 \theta}{2} \right)
\left( \vert \uparrow \rangle_1 \langle \uparrow \vert_1 
+ \vert \downarrow \rangle_1 \langle \downarrow \vert_1 \right) +
\left( \frac{\sin^2 \theta}{2} \right)
\left( \vert 0 \rangle_1 \langle 0 \vert_1 
+ \vert \uparrow \downarrow \rangle_1 \langle \uparrow \downarrow \vert_1 
\right), 
\label{eq:J7}\\
&=& \left\{ \begin{array}{ll}
\frac{1}{4}\left(\vert 0 \rangle_1 \langle 0 \vert_1
+ \vert \uparrow \rangle_1 \langle \uparrow \vert_1 
+ \vert \downarrow \rangle_1 \langle \downarrow \vert_1  
+ \vert \uparrow \downarrow \rangle_1 \langle \uparrow \downarrow \vert_1
\right)
\quad & \mbox{for $U/t \rightarrow 0$}, \\
\frac{1}{2} \left( \vert \uparrow \rangle_1 \langle \uparrow \vert_1 
+ \vert \downarrow \rangle_1 \langle \downarrow \vert_1 \right)
\quad & \mbox{for $U/t \rightarrow \infty$}. 
\end{array} \right.
\end{eqnarray}
By using Eqs. (\ref{eq:J5}), (\ref{eq:D5}), (\ref{eq:J6}) and (\ref{eq:J7}), 
we obtain the conditional entropy given by
\begin{eqnarray}
R_q &=& \frac{k_B}{(1-q)} 
\left[\frac{2^{q-1}}{(\cos^{2q} \theta +\sin^{2q} \theta)}-1 \right],
\label{eq:J8}\\
&=& \left\{ \begin{array}{ll}
\frac{k_B}{(1-q)} \left(4^{q-1}-1 \right)
\quad & \mbox{for $U/t \rightarrow 0$}, \\
\frac{k_B}{(1-q)} \left(2^{q-1}-1 \right)
\quad & \mbox{for $U/t \rightarrow \infty$}.
\end{array} \right.
\end{eqnarray}
  
For $q=1.0$ with $t/U \ll 1.0$ and $V_1=V_2=B=0$,
calculations using Eqs. (\ref{eq:K1}) and (\ref{eq:K2}) yield
\begin{eqnarray}
\frac{g_1}{X_1} &=& \frac{g_4}{X_1}
= \frac{2 t^2}{U^2}\left( 1- 3 e^{-4 \beta t^2/U}\right), \\
\frac{g_2}{X_1} &=& \frac{g_3}{X_1}
= \left( 1-\frac{2 t^2}{U^2}\right) \left( 1- 3 e^{-4 \beta t^2/U}\right)
+ 3 e^{-4 \beta t^2/U},
\end{eqnarray}
from which the conditional entropy is given by
\begin{eqnarray}
R_1 &=& S_1(1,2)-S_1(1),
\label{eq:J9} 
\end{eqnarray}
with
\begin{eqnarray}
S_1(1,2) &=& 3 \:k_B \left (1+\frac{4 \beta t^2}{U} \right) e^{-4 \beta t^2/U}, 
\label{eq:J10}\\
S_1(1) &=& k_B \left( \ln 2+\left( \frac{4t^2}{U^2} \right)
\left[1- \ln\left( \frac{4t^2}{U^2} \right) 
-6 \:e^{-4 \beta t^2/U} \right] \right).
\label{eq:J11}
\end{eqnarray}
At $T=0.0$, Eqs. (\ref{eq:J9})-(\ref{eq:J11})
yield $R_1=-k_B \ln 2$ where the negative $R_1$
expresses the quantum correlation \cite{Cerf97,Cerf99}. 
With raising the temperature, the conditional entropy is increased
and changes its sign from negative to positive because of a contribution
of the classical correlation, as will be shown is Sec. 3C.

\section{Model calculations}

\subsection{Auto- and pair correlations}

We have performed numerical calculations,
solving self-consistent equations for $E_q$ and $X_q$ given by 
Eqs. (\ref{eq:D1})-(\ref{eq:D3}), 
by using the Newton-Raphson method with
$V_1=V_2=B=0$ otherwise noticed.
Figure \ref{fig1} shows the temperature dependences of
various correlations given by 
Eqs. (\ref{eq:E13})-(\ref{eq:E15}) and (\ref{eq:F2})
for $U/t=5.0$ with $q=0.6$ and 1.0.
With increasing the temperature, 
$\sum_{\sigma} \langle  n_{1\sigma} n_{2 \sigma}\rangle_q$
is increased while both
$ \sum_{\sigma} \langle n_{1\sigma} n_{2 -\sigma}\rangle_q$ and
$\sum_{\sigma} \langle a^{\dagger}_{1\sigma} a_{2\sigma}
+ a^{\dagger}_{2\sigma} a_{1\sigma} \rangle_q$ 
are decreased. In contrast,
$ \sum_{j=1}^2 \langle n_{j\uparrow} n_{j \downarrow}\rangle_q$ 
is almost temperature independent. 
Temperature dependences of correlations for $q=0.6$
are less significant than those for $q=1.0$. 
These $q$ and $T$ dependences of correlations shown in Fig. \ref{fig1}
reflect on those of $O_q$, $L_q$, $\Gamma_q$ and $R_q$,
as will be shown in the following.

Figures \ref{fig2} (a) and (b) show the temperature dependence of 
the auto-correlation ($O_q$) for $U/t=0.0$ and 5.0, respectively,
with $q=0.6$, 0.8, 1.0 and 1.2. The magnitude of spin correlation is
given by $\langle {\bf s}_1 \cdot {\bf s}_2 \rangle_q=(3/4)O_q$.
At $T=0.0$, $O_q$ is 0.5 and 0.89 for $U/t=0.0$ and 5.0, respectively,
independently of $q$.
$O_q$ for $U/t=0.0$ is increased with increasing 
the temperature.
For $U/t=5.0$, $O_q$ is once increased 
with raising $T$, but it is decreased
at higher temperatures after showing the maximum.

Temperature dependences of pair correlation ($- L_q$)
for $U/t=0.0$ and 5.0 are shown in Figs. \ref{fig2}(c) and (d),
respectively, with $q=0.6$, 0.8, 1.0 and 1.2.
At $T=0.0$, we obtain $L_q=- 0.5$ and $- 0.89$ for $U/t=0.0$ and 5.0, respectively,
independently of $q$:
the negative sign of $L_q$ stands for antiferromagnetic correlations
for adopted parameters of $V_1=V_2=B=0.0$.
When the temperature is increased, magnitude of $L_q$ is monotonously decreased.
We note that $- L_q$ for $q < 1.0$ is larger than that for $q=1.0$ at $k_B T/t \gtrsim 0.3$,
which expresses the intrigue effect of the nonextensivity on the pair correlation. 

\subsection{Concurrence}

Figures \ref{fig2}(e) and (f) show
temperature dependences of the concurrence ($\Gamma_q$) for $U/t=0.0$ and 5.0, 
respectively, with $q=0.6$, 0.8, 1.0 and 1.2.
At $T=0.0$, $\Gamma_q=1.0$ and $0.63$ for $U/t=1.0$ and $5.0$, respectively,
independently of $q$. 
With increasing the temperature, 
$\Gamma_q$ is more slowly decreased for smaller $q$.
$\Gamma_q$ vanishes above the
threshold temperature $T_{\Gamma}$ which is implicitly determined by
\begin{eqnarray}
&&\left( \frac{4t}{D} \right) 
\vert \:{\rm Exp}_q[-\beta_{th}(\epsilon_1-E_q)]
-{\rm Exp}_q[-\beta_{th}(\epsilon_2-E_q)] \: \vert \nonumber \\
&&\;\;-{\rm Exp}_q[-\beta_{th}(\epsilon_5-E_q)]
-{\rm Exp}_q[-\beta_{th}(\epsilon_6-E_q)]=0,
\end{eqnarray}
with $\beta_{th}=1/k_B T_{\Gamma}$ and $E_q= E_q(T_{\Gamma})$.
Figures \ref{fig2}(e) and (f) show that with decreasing $q$ below unity, 
the threshold temperature $T_{\Gamma}$ is increased.
This is more clearly seen in Fig. \ref{fig3} where the solid curve
shows the $q$ dependence of $T_{\Gamma}$ for $U/t=5.0$
(the dashed curve will be explained below).

\subsection{Conditional entropy}

Figures \ref{fig2}(g) and (h) show temperature dependences 
of the conditional entropy ($R_q$) for $U/t=0.0$ and 5.0, respectively,
with various $q$ values.
$R_q$ is negative at lows temperature which expresses
the quantum entanglement \cite{Cerf97}, and it becomes positive 
at higher temperatures.
The threshold temperature $T_{R}$ at which $R_q$ changes its sign is
implicitly expressed by
\begin{eqnarray}
c_q(T_{R}) &=& d_q(T_{R}).
\end{eqnarray}
The $q$ dependence of $T_R$ is plotted by the dashed curve in Fig. 3,
which shows that $T_R$ is increased with decreasing $q$ below unity.
$T_{R}$ is correlated with $T_{\Gamma}$ as shown in the inset of Fig. 4, 
although $T_{R}$ does not agree with $T_{\Gamma}$.

We note in Figs. \ref{fig2}(a)-(h) that temperature dependences
of $O_q$, $L_q$, $\Gamma_q$ and $R_q$ become more significant
with increasing $q$, which is consistent with the more significant
temperature dependences in the specific heat ($C_q$) and 
susceptibility ($\chi_q$) shown in Fig. \ref{fig6}.

\section{Discussion}

\subsection{Relations among $L_q$, $\Gamma_q$ and $R_q$}
It is interesting to investigate the relations among $L_q$, $\Gamma_q$ and $R_q$.
In Fig. \ref{fig4}(a), $\Gamma_q$ for $U/t=0.0$ and 5.0 with $q=0.6$ and 1.0 are
plotted as a function of $-L_q$, which
shows a linear relation: $\Gamma_q \simeq a (-L_q) -b$ ($a, b>0$).
This linear relation between $\Gamma_q$ and $-L_q$ shown in Fig. \ref{fig4}(a)
is realized in the parametric
plot of $-L_q(T)$ versus $\Gamma_q(T)$
with fixed model parameters. However, it does not 
hold between $-L_q$ and $\Gamma_q$ when the model parameters are changed.
This fact is easily realized when we compared Eqs. (\ref{eq:E17}) 
with Eq. (\ref{eq:F7}), which shows that
with increasing $U$, $\vert L_q \vert$ is increased but $\Gamma_q$ is decreased.
In Fig. \ref{fig4}(b),
$\Gamma_q$ for $U/t=0.0$ and 5.0 with $q=0.6$ and 1.0 
is plotted as a function of $-R_q$, which 
shows the correlation between $\Gamma_q$ and $R_q$.
We note in Figs. \ref{fig4}(a) and (b) that $L_q$, $\Gamma_q$ and $R_q$
are correlated although the precise relations among them are not clear.


\subsection{Effect of magnetic field and interatomic interactions}

We have so far assumed $V_1=V_2=B=0.0$ for which the lowest eigenvalue 
of $\epsilon_1$ leads to the singlet ground state. 
If $V_1$, $V_2$ and/or $B$ are, however, introduced, the triplet state
may be the ground state. 
The critical condition for the singlet-triplet transition
is given by $\epsilon_1 = \epsilon_5$, {\it i.e.,}
\begin{eqnarray}
\mu_B B &=& \frac{1}{4}\left(2V_1 -V_2 -U + \sqrt{(U-V_2)^2+16 t^2} \right).
\label{eq:G1}
\end{eqnarray}


Figure \ref{fig5} shows temperature dependences of $\Gamma_q$
for $q=1.0$ (dashed curves) and $q=0.6$ (solid curves)
when $B$ is changed with $U/t=5.0$ and $V_1=V_2=0.0$, for which
Eq. (\ref{eq:G1}) yields the critical field given by
$\mu_B B_c/t=0.351$. The triplet state becomes the ground state
for $B > B_c$, where the pair correlation and
marginal entropy are positive ($L_q > 0$, $R_q > 0$) 
and the concurrence vanishes ($\Gamma_q = 0$).

Similarly, when we introduce $V_1$ and/or $V_2$ which satisfy
Eq. (\ref{eq:G1}), the triplet state becomes the ground state even if $B=0$.
In the triplet state, we obtain $L_q > 0$, $R_q > 0$ and $\Gamma_q=0$. 
The effect of interatomic interactions
on energy, entropy and specific heat of the Hubbard dimer
in the singlet state has been investigated within the NES \cite{Navarro09}. 

\subsection{Physical meaning of the entropic index}
The entropic index $q$ is conventionally regarded as a parameter which
is determined by a fitting of the $q$-exponential distribution to
experimental data except for some cases where $q$ may be determined
in a microscopic way \cite{Tsallis04}.
We will briefly discuss the physical meaning of the entropic index
in a small system coupled to finite bath
for which $q$ is theoretically derived with the use of
the MCA \cite{Plastino94,Almeida01,Potiguar03,Aringazin03,Johal03} 
and SSA \cite{Wilk00,Beck02,Cohen04}.

\noindent
{\it (1) MCA}

\noindent
We consider a microcanonical ensemble consisting of a {\it system}
and a {\it bath} with energies of $E_S$ and $E_B$,
respectively ($E=E_S+E_B$ is conserved).
Available states for the system with the energy 
between $E_S$ and $E_S+\Delta E_S$ are given by \cite{Plastino94,Potiguar03}
\begin{eqnarray}
p(E_S)\:\Delta E_S &=& \frac{\Omega_1(E_S) \Omega_2(E-E_S)}
{\Omega_{1+2}(E)} \:\Delta E_S,
\label{eq:G16}
\end{eqnarray}
where $\Omega_{\eta}(E)$ denotes the structure function expressing
the number of states with energy $E$ in $\eta$ (=S, B and S+B).
We assume that the structure function is given by \cite{Plastino94,Potiguar03}
\begin{eqnarray}
\Omega_{\eta}(E) &=& K m_{\eta} \:E^{m_{\eta}-1},
\label{eq:G7}
\end{eqnarray}
where  $K$ is a constant and $m_{\eta}$ the degree of freedom of variables
in $\eta$.
Eq. (\ref{eq:G7}) is justified for $d$-dimensional $N$-body ideal gases 
and harmonic oscillators, for which $m=dN/2$ and $dN$, respectively.
For $E_S \ll E_B$ and $m_S \ll m_B$, Eqs. (\ref{eq:G16}) and (\ref{eq:G7}) 
lead to the PDF given by \cite{Plastino94,Potiguar03}
\begin{eqnarray}
p(E_S) &\propto& \left( 1- \frac{E_S}{E}\right)^{m_B}, 
\label{eq:G8c} \\
& \propto & {\rm Exp}_q(- q \beta E_S),
\label{eq:G12}
\end{eqnarray}
with
\begin{eqnarray}
q &=& 1+ \frac{1}{m_B},
\label{eq:G13} \\
\beta &=& \frac{1}{(q-1)E},
\label{eq:G14}
\end{eqnarray}
where ${\rm Exp}_q(x)$ denotes the $q$-exponential function
given by Eq. (\ref{eq:C6}).
Eq. (\ref{eq:G12}) corresponds to the PDF obtained in the normal average.
In the case of $m_B \rightarrow \infty$, Eq. (\ref{eq:G8c}) reduces to
\begin{eqnarray}
p(E_S) &\propto& e^{-\beta E_S},
\label{eq:G8b}
\end{eqnarray}
with
\begin{eqnarray}
\beta &=& \frac{m_B}{E} = \frac{1}{k_B T}.
\end{eqnarray}
The specific heat of the bath is shown to be given by \cite{Almeida01}
\begin{eqnarray}
C_v &=& \frac{d E_B}{d T} \propto \frac{1}{q-1}.
\label{eq:G10}
\end{eqnarray}
Eqs. (\ref{eq:G12}), (\ref{eq:G13}), (\ref{eq:G8b})  
and (\ref{eq:G10}) imply that for finite bath,
the PDF is given by the $q$-exponential function
whereas for infinite bath ($C_v = \infty$), 
the PDF is given by the exponential function. 
A similar analysis has been made within the microcanonical approach in Refs. 
\cite{Aringazin03,Johal03}.

\vspace{0.5cm}
\noindent
{\it (2) SSA} 

\noindent
In the superstatistics \cite{Wilk00,Beck02,Cohen04}, it is assumed that 
properties of a given system
may be expressed by a superposition over the spatially and/or
temporarily fluctuating intensive parameter ({\it i.e.,} the inverse temperature) 
\cite{Wilk00,Beck02,Cohen04}.
Since the PDF of the equilibrium state $i$
with the inverse temperature $\beta$ ($=1/k_B T$) is given by
$e^{-\beta \epsilon_i}/Z_1(\beta)$,
the PDF averaged over fluctuating $\beta$-variable is assumed to be given by 
\cite{Wilk00,Beck02,Cohen04}
\begin{eqnarray}
p_i &=& \int_{0}^{\infty}\: \frac{e^{-\beta \epsilon_i}}{Z_1(\beta)} 
\:f(\beta)\:d \beta,
\label{eq:G2}
\end{eqnarray}
with
\begin{eqnarray}
f(\beta) &=& 
\frac{1}{\Gamma(n/2)}\left(\frac{n}{2\beta_0} \right)^{n/2}
\beta^{n/2-1} e^{-n \beta/2 \beta_0}.
\label{eq:G3b}
\end{eqnarray}
Here $\Gamma(x)$ is the gamma function and
$f(\beta)$ denotes the $\chi^2$-distribution with rank $n$ which
expresses the distribution of sum of squares of $n$ random normal variables
with zero mean and unit variance \cite{Beck02}.
Average and variance of $\beta$ 
are given by $ \langle \beta \rangle_{\beta}=\beta_0 $
and $(\langle \beta^2 \rangle_{\beta}-\beta_0^2)/\beta_0^2=2/n$.
When adopting the type-A superstatistics in which the $\beta$ dependence
in $Z_1(\beta)$ is neglected \cite{Cohen04}, we obtain (with $n=n_S$),
\begin{eqnarray}
p_i &\propto& \left(1+\frac{2 \beta_0}{n_S}\epsilon_i \right)^{-n_S/2}, 
\label{eq:G5} 
\end{eqnarray}
which is rewritten as
\begin{eqnarray}
p_i 
&\propto& {\rm Exp}_q\left(-q \beta_0 \epsilon_i \right),
\label{eq:G3}
\end{eqnarray}
with
\begin{eqnarray}
q &=& 1-\frac{2}{n_S}.
\label{eq:G4}
\end{eqnarray}
Eq. (\ref{eq:G3}) is in conformity with the normal-average PDF. 

It has been shown by a detailed microscopic calculation 
that the distribution of the inverse temperature
of a system containing independent $n$ particles
coupled to a bath characterized by a fixed inverse temperature of
$\bar{\beta}$, is given by \cite{Touch02}
\begin{eqnarray}
f_T(\beta) &=& \frac{\bar{\beta}}{\Gamma(n/2)}
\left( \frac{n \bar{\beta}}{2 }\right)^{n/2}
\beta^{-n/2-2}\:e^{-n \bar{\beta}/2 \beta}.
\label{eq:L1}
\end{eqnarray}
Eq. (\ref{eq:L1}) expresses the inverse-gamma
distribution, and its profile is similar to that of
the gamma-distribution given by Eq. (\ref{eq:G3b})
for large $n$  \cite{Touch02}. Unfortunately we cannot obtain an analytic
expression for the PDF averaged over $f_T(\beta)$
by using Eq. (\ref{eq:G2}) with $f(\beta) \rightarrow f_T(\beta)$.
Nevertheless the calculation of Ref.  \cite{Touch02} 
justifies the concept of the superstatistics.

\section{Conclusion}

We have calculated various quantities of quantum entanglement
such as auto- ($O_q$) and pair correlations ($L_q$), concurrence ($\Gamma_q$) 
and conditional entropy ($R_q$) of the half-filled Hubbard dimer
as functions of the entropic index and 
the temperature within the framework of the NES \cite{Tsallis88}.
It has been shown that the properties of $O_q$, $L_q$, $\Gamma_q$ and $R_q$
are considerably modified by the nonextensivity.
In particular, for $q < 1.0$, the thermal entanglement
my be survive at higher temperatures than that for $q=1.0$, 
because the threshold temperature 
where $\Gamma_q$ vanishes ($T_{\Gamma}$) or $R_q$ changes 
its sign ($T_R$) is more raised for a larger $(1-q)$ (Fig. 3).
The three measures of $L_q$, $\Gamma_q$ and $R_q$ for thermal entanglement
are correlated each other although the precise relations among them are not clear.

The NES has a wider applicability than the BGS, which corresponds to 
the $q=1.0$ limit of the NES.
We note that the PDF in the MCA given by Eq. (\ref{eq:G12}) is equivalent to
that in the SSA given by Eqs. (\ref{eq:G3}).
There is, however, distinct differences in their expressions of $q$ 
[Eqs. (\ref{eq:G13}) and (\ref{eq:G4})] \cite{Note2,Note4,Note3}:
\begin{eqnarray}
q &= & \left\{ \begin{array}{ll}
1+\frac{1}{m_B} \geq 1.0
\quad & \mbox{in the MCA}, \\ 
1-\frac{2}{n_S} \leq 1.0
\quad & \mbox{in the SSA}.
\end{array} \right.
\label{eq:L10}
\end{eqnarray}
The entropic index in the MCA is expressed in terms of the bulk parameter ($m_B$)
while that in the SSA is given in terms of the system one ($n_S$).
Furthermore the conceivable value of $q$ in the MCA is different from that 
in the SSA. In this respect we have not obtained a unified physical interpretation of
the entropic index at the moment. Nevertheless
Eq. (\ref{eq:L10}) shows that the entropic index $q$ may be related
with the size of the system and/or bath and that
the nonextensivity may be realized in such a small-scale system.
It might be interesting to perform experiments by changing the
size of the system and/or bath, in order to examine the possibility
that the nonextensivity reflects on the thermal entanglement
of two-qubit Hubbard dimer. 
Such experimental studies might clarify the role of
the nonextensivity in small systems and 
provide valuable insight on the validity of the MCA and SSA.


\section*{Acknowledgements}
This work is partly supported by
a Grant-in-Aid for Scientific Research from the Japanese 
Ministry of Education, Culture, Sports, Science and Technology.  

\vspace{0.5cm}
\appendix*

\section{A. Thermodynamical quantities of the Hubbard dimers with the original MEM}

\subsection{Energy and entropy}

When the energy $E_q$ and partition function $X_q$ are obtained by solving 
Eqs. (\ref{eq:D1})-(\ref{eq:D3}) for given $q$ and $\beta$, 
the Tsallis entropy given by Eq. (\ref{eq:A1}) may be calculated by
\begin{eqnarray}
S_q &=& k_B \left( \frac{c_q-1}{1-q} \right),
\label{eq:D4}
\end{eqnarray}
with
\begin{eqnarray}
c_q&=& Tr \:(\hat{\rho}_q^q)
= \frac{1}{X_q^q}\sum_i\: w_i^q = X_q^{1-q}.
\end{eqnarray}

\subsection{Specific heat}
The specific heat is given by \cite{Hasegawa04,Hasegawa05}
\begin{eqnarray}
C_q &=& \left( \frac{d \beta}{dT} \right)
\left( \frac{\partial E_q}{\partial \beta} \right),
\end{eqnarray}
where $\partial E_q/\partial \beta$ may be determined by
simultaneous equations given by
\begin{eqnarray}
\frac{\partial E_q}{\partial \beta} 
&=& a_{11} \frac{\partial E_q}{\partial \beta} 
+ a_{12} \frac{\partial X_q}{\partial \beta} + b_1, 
\label{eq:D6}\\
\frac{\partial X_q}{\partial \beta} 
&=& a_{21} \frac{\partial E_q}{\partial \beta}
+ a_{22} \frac{\partial X_q}{\partial \beta}+b_2,
\label{eq:D7} 
\end{eqnarray}
with
\begin{eqnarray}
a_{11} &=& \frac{\beta}{q X_q} \sum_i w_i^{2-q}\:\epsilon_i, 
\;\;\;\; a_{12} = - \frac{E_q}{X_q}, 
\;\;\; a_{21} = \beta X_q, \;\;a_{22} = 0, 
\nonumber 
\\
b_1 &=& -\frac{1}{q X_q} \sum_i w_i^{2-q} \epsilon_i (\epsilon_i-E_q),
\;\; b_2 = 0.
\label{eq:E7b}
\end{eqnarray}
Eqs. (\ref{eq:D6})-(\ref{eq:E7b}) are derived from 
Eqs. (\ref{eq:D1})-(\ref{eq:D3}).
Solving Eqs. (\ref{eq:D6})-(\ref{eq:E7b}) for $\partial E_q/\partial \beta$, 
we obtain
\begin{eqnarray}
C_q &=& - \left( \frac{1}{k_B T^2} \right)
\frac{b_1}{1-a_{11}-a_{12}a_{21}}.
\end{eqnarray}
In the limit of $q=1.0$, $C_q$ becomes
\begin{eqnarray}
C_1 &=&  \left( \frac{1}{k_B T^2} \right)(\langle \epsilon_i^2 \rangle_1
- \langle \epsilon_i \rangle_1^2).
\end{eqnarray}

\subsection{Susceptibility}
The paramagnetic spin susceptibility $\chi_q$ is given by
\cite{Hasegawa04,Hasegawa05}
\begin{eqnarray}
\chi_q &=& -E_q^{(2)}
+\frac{1}{\beta X_q} X_q^{(2)},
\label{eq:D11}
\end{eqnarray}
where $E_q^{(2)}=\partial^2 E_q/\partial B^2\vert_{B=0}$ and
$X_q^{(2)}=\partial^2 X_q/\partial B^2\vert_{B=0}$.
From Eqs. (\ref{eq:D1})-(\ref{eq:D3}), we obtain
the simultaneous equations for 
$E_q^{(2)}=\partial^2 E_q/\partial B^2\vert_{B=0}$ and
$X_q^{(2)}=\partial^2 X_q/\partial B^2\vert_{B=0}$,
\begin{eqnarray}
E_q^{(2)} &=& a_{11} E_q^{(2)}+ a_{12} X_q^{(2)}+f_1,
\label{eq:D12}\\
X_q^{(2)} &=& a_{21} E_q^{(2)}+ a_{22} X_q^{(2)}+f_2,
\label{eq:D13}
\end{eqnarray}
with
\begin{eqnarray}
f_2 &=& \frac{\beta^2}{q} \:\sum_i \:w_i^{2-q} \mu_i^2.
\label{eq:D14}
\end{eqnarray}
From Eqs. (\ref{eq:D11})-(\ref{eq:D14}), we obtain 
\begin{eqnarray}
\chi_q &=& \frac{f_2}{a_{21}}
= \frac{\beta}{q X_q} \sum_i w_i^{2-q} \mu_i^2,
\label{eq:D15}
\end{eqnarray}
which does not include $f_1$.
In the limit of $q=1.0$, the spin susceptibility is given by
\begin{eqnarray}
\chi_1 &=& \frac{\beta}{Z_1} \sum_i e^{-\beta \epsilon_i}\:\mu_i^2.
\end{eqnarray}

\subsection{Model calculations}

Fig. \ref{fig6} shows temperature dependences of the calculated
entropy, specific heat and susceptibility.
Temperature dependences of the entropy $S_q$ for $U/t=0.0$ and 5.0 
are plotted in Figs. 6(a) and (d), respectively, with
$q=0.6$, 0.8, 1.0 and 1.2. 
For a larger $U$ value, $S_q$ is more quickly increased
at low temperatures. This is because the energy difference between the
ground state ($\epsilon_1$) and the first-excited state ($\epsilon_4$)
becomes smaller when the strength of $U$ is more increased. 
With increasing $q$, the saturation value of $S_q$ at higher
temperatures becomes smaller.
Temperature dependences of specific heat $C_q$ for $U/t=0.0$ and 5.0
are plotted in Figs. 6(b) and (e), respectively,
for various $q$ values.
Figs. 6(c) and (f) show temperature dependences of 
the susceptibility $\chi_q$ for $U/t=0.0$ and 5.0, respectively,
with $q=0.6$, 0.8, 1.0 and 1.2.
We note that temperature dependences of $C_q$ and $\chi_q$
at low temperatures for $q=1.2$ are more significant than those of $q=1.0$
whereas those for $q=0.8$ is less significant than those of $q=1.0$:
temperature dependences of $C_q$ and $\chi_q$ become
more significant with increasing $q$.
In contrast, $S_q$ is increased with increasing $q$.
These behaviors are understood as follows. The $q$- and $T$-dependent
thermodynamical quantity $Q_q(T)$ may be expand at $q=1.0$,
\begin{eqnarray}
Q_q(T) &=& \sum_{k=0}^{\infty} \frac{(q-1)^k}{k!} Q_q^{(k)}(T), \\
&\simeq& Q_1(T)+ (q-1) Q_1^{(1)}(T) + \cdot\cdot, 
\end{eqnarray}
where $Q_1^{(k)}(T)=\partial^k Q_q(T)/\partial q^k \vert_{q=1}$.
Actual analytical evaluation of $Q_1^{(1)}(T)$ is tedious because
it involves self-consistent calculations as discussed in preceding subsections.
Our model calculations show that $Q_1^{(1)}(T) > 0$ for $C_q$ and $\chi_q$
whereas $Q_1^{(1)}(T) < 0$ for $S_q$ at low temperatures.
The characteristic temperature dependences in thermodynamical quantities
depend on the entropic index $q$.
When comparing these results with the counterparts 
obtained in our previous study \cite{Hasegawa04} 
using the normalized MEM with $q$-average \cite{Tsallis98}, 
we realize that both results approximately have 
the $q \leftrightarrow 1/q $ symmetry:
for example, results of $q=0.6$ in Fig. \ref{fig6}
are similar to those of $q=1.5$ 
in Figs. 2, 3 and 4 of Ref. \cite{Hasegawa04,Note1}.

\section{B. Heisenberg dimers with the original MEM}

\renewcommand{\theequation}{B\arabic{equation}}
\setcounter{equation}{0}

It is well known that the Hubbard dimer with $U/t \gg 1$ (with $V_1=V_2=0$)
is equivalent to the Heisenberg dimer with the superexchange
interaction of $J_{se} \sim - t^2/U$.
It is worthwhile to apply the NES with the original MEM \cite{Tsallis88,Abe09}
to a Heisenberg dimer given by ($s=1/2$)
\begin{eqnarray}
H &=& -J {\bf s}_1 \cdot {\bf s}_2 
- g \mu_B B (s_{1z}+s_{2z}),
\label{eq:H1}
\end{eqnarray}
where the exchange interaction $J$ is positive (negative) 
for ferromagnetic (antiferromagnetic) coupling,
$g$ (=2) denotes the g-factor, $\mu_B$ the Bohr magneton,
and $B$ an applied magnetic field.
Four eigenvalues of $H$ are given by
\begin{eqnarray}
\epsilon_i &=& \frac{3 J}{4},\;\;-\frac{J}{4}, 
\;\; -\frac{J}{4}- g \mu_B B, 
\;\; -\frac{J}{4}+ g \mu_B B 
\hspace{1cm}\mbox{for $i=1, 2, 3$ and 4},
\label{eq:H2}
\end{eqnarray}
and corresponding eigenvectors are given by
\begin{eqnarray}
\vert \Phi_1\rangle &=& \frac{1}{\sqrt{2}}
\left( \vert \uparrow \rangle_1 \vert \downarrow \rangle_2 
- \vert \downarrow \rangle_1 \vert \uparrow \rangle_2 \right),
\label{eq:H21}\\
\vert \Phi_2 \rangle &=& \frac{1}{\sqrt{2}}
\left( \vert \uparrow \rangle_1 \vert \downarrow \rangle_2
+ \vert \downarrow \rangle_1 \vert \uparrow \rangle_2 \right),
\label{eq:H22}\\
\vert \Phi_3 \rangle &=& \vert \uparrow \rangle_1 \vert \uparrow\rangle_2,\\
\vert \Phi_4 \rangle &=& \vert \downarrow \rangle_1 \vert \downarrow \rangle_2.
\end{eqnarray}
When $B=0$, $\vert \Phi_3 \rangle$ and 
$\vert \Phi_4 \rangle$ may be alternatively expressed by
\begin{eqnarray}
\vert \Phi_3 \rangle &=& \frac{1}{\sqrt{2}}
\left( \vert \uparrow \rangle_1 \vert \uparrow\rangle_2
- \vert \downarrow \rangle_1 \vert \downarrow \rangle_2
\right),
\label{eq:H23}\\
\vert \Phi_4 \rangle &=& \frac{1}{\sqrt{2}}
\left( \vert \uparrow \rangle_1 \vert \uparrow\rangle_2
+ \vert \downarrow \rangle_1 \vert \downarrow \rangle_2
\right).
\label{eq:H24}
\end{eqnarray}
Eqs. (\ref{eq:H21}), (\ref{eq:H22}), (\ref{eq:H23})
and (\ref{eq:H24}) form the Bell basis.

In the BGS the partition function is given by
\begin{eqnarray}
Z_{BG} &=&Z_1 = {\rm e}^{-\frac{3 \beta J}{4}}
+{\rm e}^{\frac{\beta J}{4}}
\left[1 + 2 {\rm cosh} \left(2 \beta \mu_B B \right) \right],
\label{eq:H3}
\end{eqnarray}
from which various thermodynamical quantities are
easily calculated.
The susceptibility is, for example, given by
\begin{eqnarray}
\chi_{BG} &=& 
\left( \frac{\mu_B^2}{k_B T} \right) 
\left( \frac{8}{3+ {\rm e}^{-\beta J} } \right).
\label{eq:H4}
\end{eqnarray}

The calculation for the Heisenberg dimer within the NES goes parallel
to that for the Hubbard dimer with the use of
four eigenvalues given by Eq. (\ref{eq:H2}).
The average energy and partition function are expressed by
\begin{eqnarray}
E_q &=& \frac{1}{X_q} \sum_i w_i \:\epsilon_i,
\label{eq:H5} \\
X_q &=& \sum_i w_i, 
\label{eq:H6}
\end{eqnarray}
with
\begin{eqnarray}
w_i &=& {\rm Exp}_q[-\beta (\epsilon_i-E_q)].
\label{eq:H7}
\end{eqnarray}

The pair-correlation function ($L_q$) is defined by
\begin{eqnarray}
\left(\frac{3}{4}\right) L_q &=& \left< {\bf s}_1 \cdot {\bf s}_2 \right>_q, \\
&=& - \left< \frac{\partial H}{\partial J}\right>_q, \\
&=& \frac{1}{X_q} \sum_i w_i^q 
\: \left( - \frac{\partial \epsilon_i}{\partial J} \right),
\label{eq:H9}
\end{eqnarray}
which yields (with $B=0.0$ hereafter)
\begin{eqnarray}
L_q &=& \frac{1}{X_q} 
\{ \left({\rm Exp}_q\left[-\beta\left(-\frac{J}{4}-E_q \right) \right] \right)^q 
-\left( {\rm Exp}_q\left[-\beta\left(\frac{3J}{4}-E_q \right) \right] \right)^q \}.
\label{eq:H10}
\end{eqnarray}
For $q=1.0$ with $B=0$, Eq. (\ref{eq:H10}) reduces to
\begin{eqnarray}
L_1 &=& \frac{1}{Z_1} 
\left( e^{\frac{\beta J}{4}}-e^{-\frac{3 \beta J}{4}} \right).
\end{eqnarray}

The concurrence for $J < 0$ is given by

\begin{eqnarray}
\Gamma_q &=&
\frac{1}{X_q}
{\rm max}\left[\vert \:2\: w_1 - \sum_i \:w_i \:\vert ,0 \right].
\label{eq:H12}
\end{eqnarray}
For $q=1.0$ and $B=0$, Eq. (\ref{eq:H12}) becomes
\begin{eqnarray}
\Gamma_q &=& \frac{1}{Z_1} {\rm max}\left[\vert \:e^{-\frac{3\beta J}{4}}
-3 e^{\frac{\beta J}{4}}\:\vert,0 \right],
\end{eqnarray}
which is in agreement with the result of Ref. \cite{Arnesen01}.

By using the marginal density matrix given by 
\begin{eqnarray}
\hat{\rho}_q(1) &=& \frac{1}{2}
\left( \vert \uparrow \rangle_1 \langle\uparrow \vert_1 
+ \vert \downarrow \rangle_1 \langle \downarrow \vert_1 \right),
\end{eqnarray}
we obtain the conditional entropy \cite{Abe01},
\begin{eqnarray}
R_q &=& k_B \left[\frac{(2^{q-1} X_q^{1-q}-1)}{(1-q)} \right].
\end{eqnarray}



\newpage
\begin{figure}
\begin{center}
\end{center}
\caption{
(Color online)
(a) Temperature dependences of various correlations for $U/t=5.0$
with $q=0.6$ (solid curves) and $1.0$ (dashed curves).
}
\label{fig1}
\end{figure}

\begin{figure}
\begin{center}
\end{center}
\caption{
(Color online)
Temperature dependences of auto-correlation ($O_q$) 
for (a) $U/t=0.0$ and (b) $U/t=5.0$,
pair correlation ($- L_q$) for (c) $U/t=0.0$, (d) $U/t=5.0$, 
concurrence ($\Gamma_q$) for (e) $U/t=0.0$ and (f) $U/t=5.0$,
and the conditional entropy ($R_q$) for (g) $U/t=0.0$ and (h) $U/t=5.0$ with
$q=0.6$ (solid curves), 0.8 (dotted curves), 1.0 (dashed curves)
and 1.2 (chain curves).
}
\label{fig2}
\end{figure}

\begin{figure}
\begin{center}
\end{center}
\caption{
(Color online)
$q$ dependences of threshold temperatures of $T_{\Gamma}$ (the solid curve)
and $T_R$ (the dashed curve) for $U/t=5.0$,
the inset showing $T_{R}$ against $T_{\Gamma}$.
}
\label{fig3}
\end{figure}

\begin{figure}
\begin{center}
\end{center}
\caption{
(Color online)
(a) $\Gamma_q$ as a function of $-L_q$ and 
(b) $\Gamma_q$ as a function of $-R_q$
with $q=0.6$ (solid curves) and $1.0$ (dashed curves),
the result of $U/t=5.0$ in (a) for $q=0.6$ being indistinguishable from 
that for $q=1.0$.
}
\label{fig4}
\end{figure}

\begin{figure}
\begin{center}
\end{center}
\caption{
(Color online)
Temperature dependences of the pair correlation $L_q$ for various $B$ values 
for $U/t=5.0$ with $q=0.6$ (solid curves) and $1.0$ (dashed curves).
}
\label{fig5}
\end{figure}

\begin{figure}
\begin{center}
\end{center}
\caption{
(Color online)
The temperature dependences of (a) $S_q$, (b) $C_q$ and (c) $\chi_q$ for $U/t=0$,
and (d) $S_q$, (e) $C_q$ and (f) $\chi_q$ for $U/t=5$ with
$q=0.6$ (solid curves), $0.8$ (dotted curves), $1.0$ (dashed curves)
and $1.2$ (chain curves). 
}
\label{fig6}
\end{figure}

\end{document}